\documentclass[hidelinks,a4paper]{jpconf}
\usepackage{graphicx}
\usepackage{iopams}
\usepackage{amsmath,amsfonts,tikz,tikz-cd,float,listings,color,pdfpages,enumitem,amssymb,hyperref,bm,amsthm}

\newcommand\ddfrac[2]{\frac{\displaystyle #1}{\displaystyle #2}}
\usepackage{fancyhdr}

\begin{document}
\title{The diffraction volume for square-shaped samples in X-ray diffraction with high spatial resolution}

\author{P~Chakrabarti\textsuperscript{1,}\textsuperscript{2} and P~Modregger\textsuperscript{1,}\textsuperscript{2}}

\address{\textsuperscript{1} Physics Department, University of Siegen, 57072 Siegen, Germany}
\address{\textsuperscript{2} Center for X-ray and Nano Science CXNS, Deutsches Elektronen-Synchrotron DESY, 22607 Hamburg, Germany}

\ead{prerana.chakrabarti@uni-siegen.de}
\begin{abstract}
X-ray diffraction with high spatial resolution is a prerequisite for the characterization of (poly)-crystalline materials on micro- or nanoscopic scales. This can be achieved by utilizing a focused X-ray beam and scanning of the sample. However, due to the penetration of the X-rays into the material, the exact location of diffraction within the sample is ambiguous. Here, we utilize numerical simulations to compute the spatially resolved diffraction volume in order to investigate these ambiguities. We demonstrate that partial depth sensitivity can be achieved by rotating the sample.
\end{abstract}

\section{Introduction}
X-ray diffraction (XRD) with high spatial resolution provides access to micro strain, residual stress or crystallite sizes on micro- or nanoscopic scales. High spatial resolution can be achieved by combining XRD with focussing optics such as compound refractive lenses (CRLs)~\cite{crl} or Kirpatrick-Baez mirrors~\cite{kbmirror} and scanning of the sample through the focus.

Examples for X-ray diffraction using high spatial resolution include the following. Analysis of local strains and grain boundaries in functional thin films with a spot size of $(100\times 100)\ \mathrm{nm}^2$ at the ID01 beamline of the European Synchrotron Radiation Facility~\cite{microdiffraction:thinfilm}. The determination of local micro-structure of martensite-retained austenite steel with a spot size of $(4\times 1)\,\mu \mathrm{m}^2$ at the 2-ID-D undulator beamline of the Advanced Photon Source~\cite{microdiffraction:cai}. The renovated high-pressure XRD setup at the BL10XU beamline of SPring-8~\cite{SPring-8:renovation} utilizes a spot size of $(1\times 1)\,\mu \mathrm{m}^2$. The design and construction of a microscopy system to accommodate multilayer Laue lenses in order to increase the imaging resolution to 10~nm utilizes a spot size of $(13\times 33)\ \mathrm{nm}^2$ at the I13 beamline of Diamond Light Soure~\cite{nanoimaging}.

\begin{figure}
\centering
\includegraphics[width=0.45\linewidth]{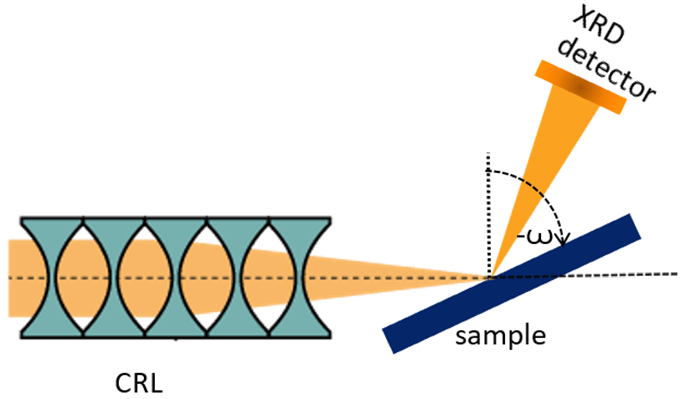}
\caption{Sketch of the setup geometry (top view).}
\label{fig:setup}
\end{figure}

Investigations of highly absorbing samples with high spatial resolution depend on the simultaneous use of high photon energies and small spot sizes. Recently, such a setup was implemented at P06~\cite{p06} of PETRA III at DESY, Hamburg. A sketch of the setup is shown in \Fref{fig:setup}. The incident photon energy of 35~keV was selected by a double crystal monochromator and CRLs were used to focus the beam down to $(2.0\times 1.2)\,\mu \mathrm{m}^2$. Samples were mounted on a novel 6 axes goniometer, which allowed for spatial scans at fixed angles. More details can be found in~\cite{hrxrd}.

We have used martensitic steel as powder-like samples providing distinct and texture-less Laue rings. Martensite (i.e., $\alpha '$-iron) is a meta-stable phase of carbon steel, which is distinguished by its micro- and nanoscopic lamellar structure~\cite{Krauss1999}. Martensitic steels provide ultimate tensile strengths beyond 2~GPa and exhibit excellent resilience in high cycle fatigue regime which renders them the material choice for suspension springs in vehicles~\cite{Tump2016}. The samples had quadratic cross-sections of size $\approx (1.0\times 1.0)\, \mathrm{mm}^2$ and the photon energy implied an absorption length of 250~$\mu$m~\cite{hrxrd}. 

Nonetheless, due to the penetration of the X-rays into the sample, the precise location of diffraction within the sample is uncertain. In the following, we describe a framework for numerical simulations of the spatially resolved diffraction volume of the sample. The principles laid out are applicable to the high resolution XRD with powder-like sample but we will use the specific experimental parameters of the setup and sample described above.

\section{Framework for numerical simulations}

\begin{figure}[H]
\centering
\includegraphics[width=0.5\linewidth]{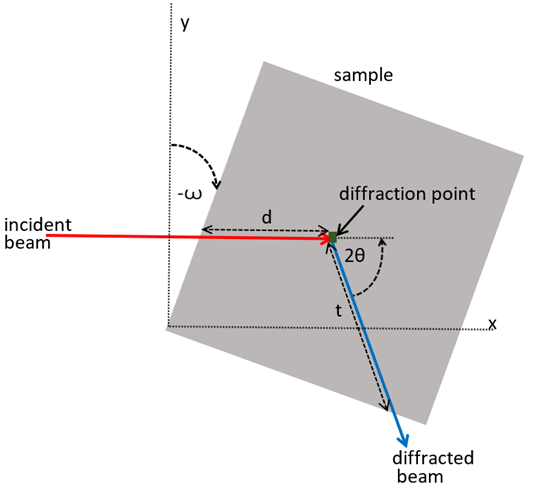}
\caption{\label{ray trace}Ray tracing quantities used to analyze diffraction volume of the sample. }
\end{figure}

We have used a ray tracing approach for the computation of the diffraction volume and \Fref{ray trace} shows a sketch with the quantities of interest. The focused incident beam penetrates the sample and diffracts with the angle $2\theta$. The diffraction point is marked by a square. Then, the diffracted beam exits the sample through one of its edges. The incident angle $\omega$ can be changed by rotating the sample. The distances between the diffraction point and the entry or exit point are $d$ and $t$, respectively. In order to compute the detectable relative intensities, Beer-Lambert law, given as 
\begin{equation}\label{beer's_law}
    I(x,y) = I_0 e^{-\mu (d(x,y)+t(x,y))},
\end{equation}
was used. $I_0$ is the incident intensity, $\mu$ is the linear attenuation coefficient, and $d+t$ is the total path length traveled by the beam in the sample. 

Such a diffraction volume would correspond to a hypothetical sample that contains homogeneously distributed and randomly oriented crystallites and, thus, provides the possibility of diffraction at all positions. Though, this does not apply to a specific sample, we use such a diffraction volume here in order to investigate the information depth on average.

For the following simulations, we used the linear attenuation coefficient of martensitic steel $\mu = 42.93\, \mathrm{cm}^{-1}$~\cite{nist}. \Tref{hkl} shows the angular positions 2$\theta$ and their corresponding (hkl) reflections of martensitic steel at 35~keV, which were used for the simulations. Here, we assume martinsite is a bcc crystal system with a lattice parameter of $a=2.866$~\AA~\cite{Kim2014}.

\begin{table}[htbp]
\caption{\label{hkl}Bragg angles 2$\theta$ of martensitic steel (bcc with lattice parameter $a=2.866$ \AA) at 35~keV for different reflections (hkl).}
\begin{center}
\begin{tabular}{llllll}
\br
2$\theta$ & 14.19$^\circ$ & 17.44$^\circ$ & 20.18$^\circ$ & 22.56$^\circ$ & 24.77$^\circ$\\
(hkl) & (200) & (211) & (220) & (310) & (222)\\
\br
\end{tabular}
\end{center}
\end{table}

\section{Results}
\begin{figure}[htbp]
\centering
\includegraphics[width=0.8\textwidth]{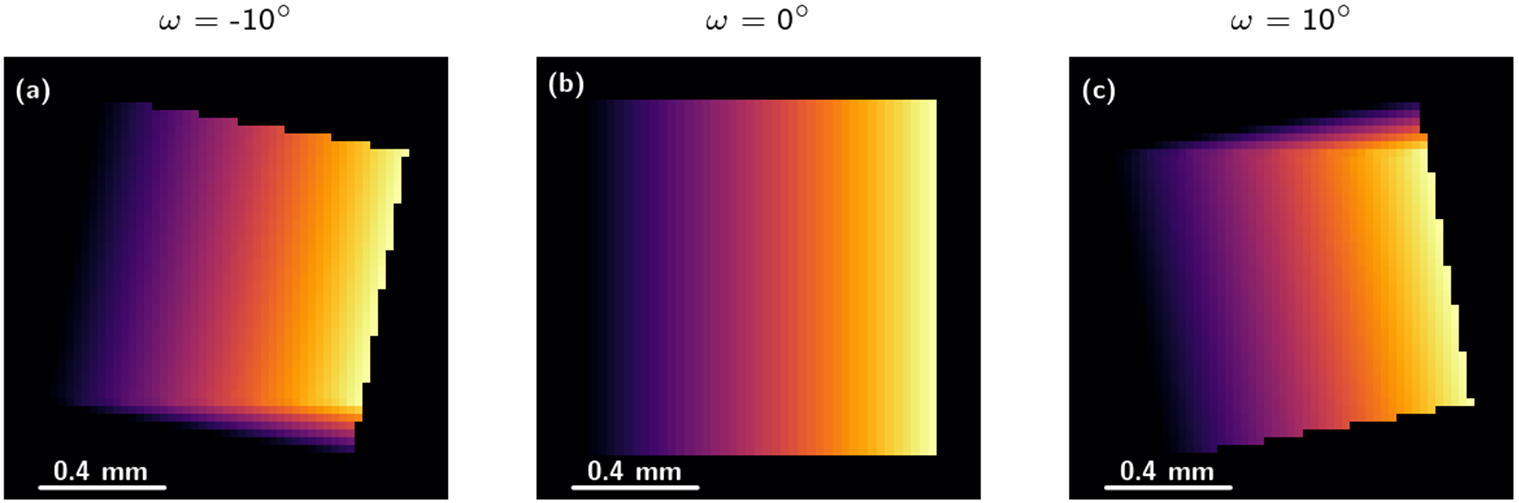}\\
\includegraphics[width=0.13\textwidth, angle =-90 ]{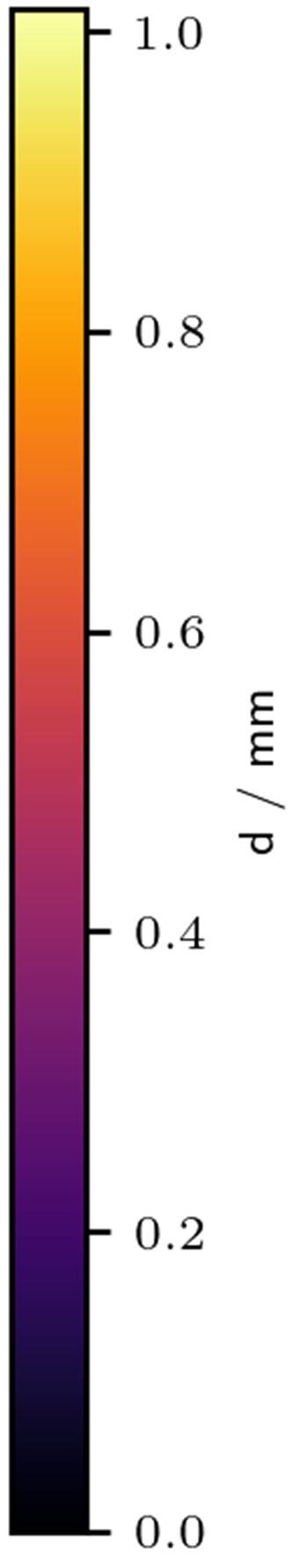}\\
\caption{\label{distance_d}Dependency of distance $d$ as a function of incident angle (a) $\omega$ = -10$^\circ$ (b) $\omega$ = 0$^\circ$ and (c) $\omega$ = 10$^\circ$.}
\end{figure}

\Fref{distance_d} demonstrates the spatially resolved dependency of the distance between entry point and diffraction point $d$ as a function of the incident angle $\omega$. Please note that $d$ does not depend on $2\theta$. As expected $d$ increases linearly with increasing penetration depth.

\begin{figure}[htbp]
\centering
\includegraphics[width=0.8\textwidth]{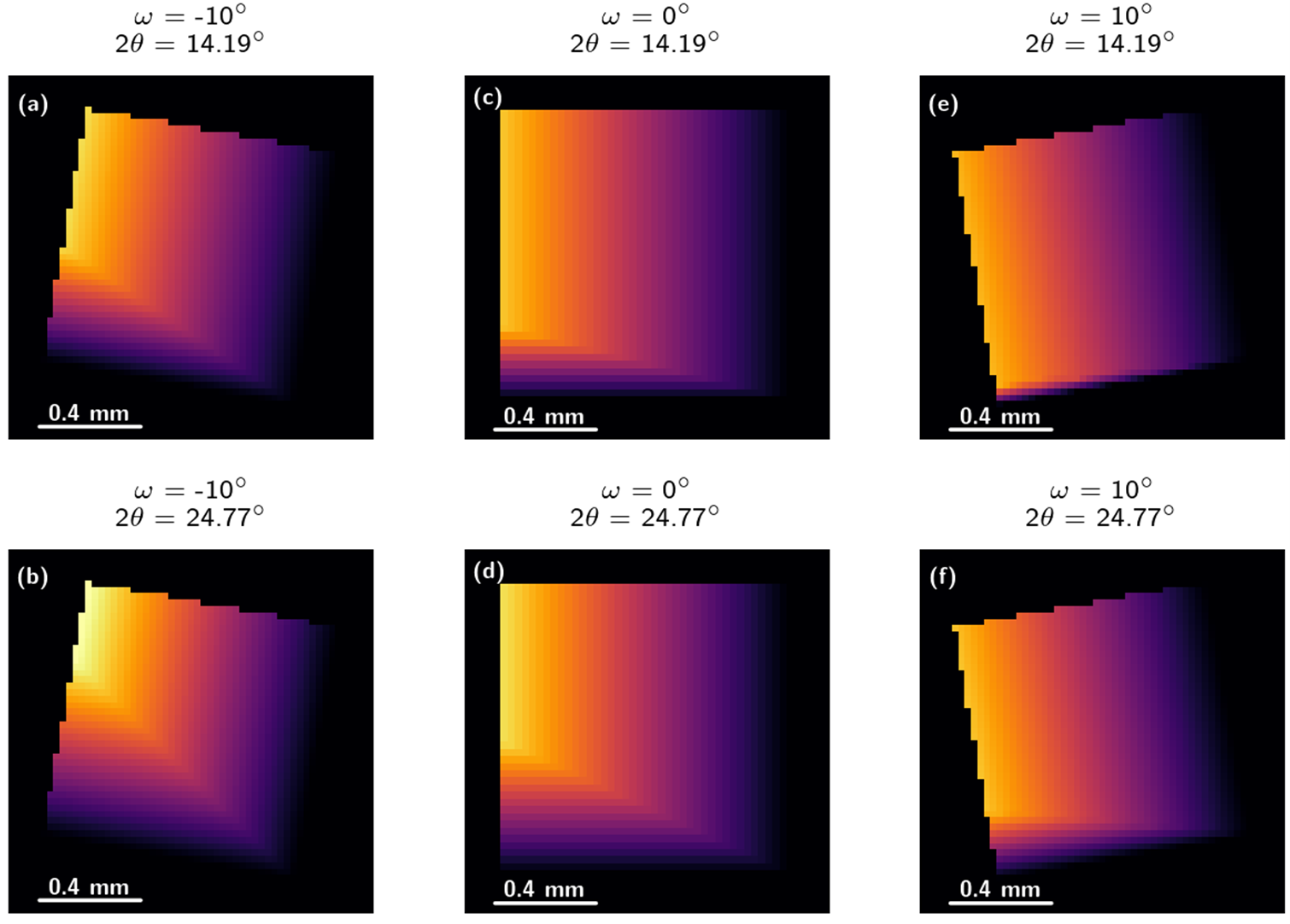}\\
\includegraphics[width=0.13\textwidth, angle =-90 ]{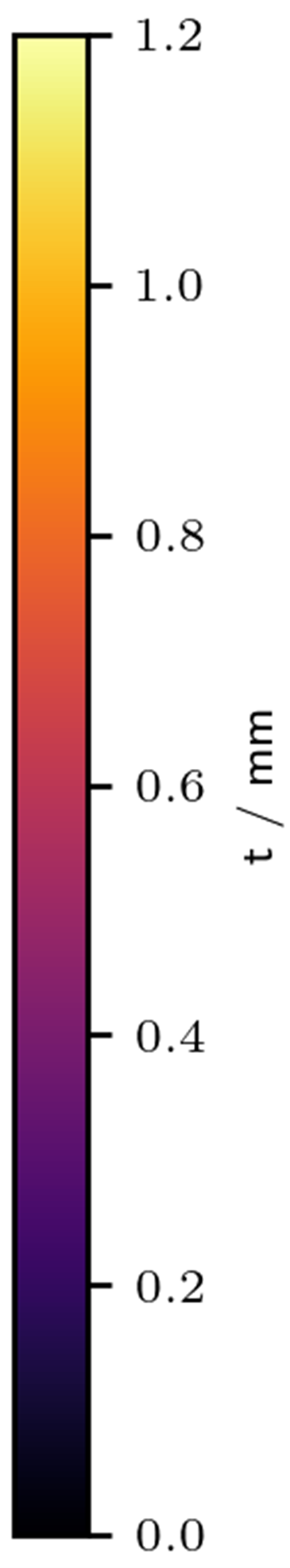}\\
\caption{\label{distance_t}Dependency of the distance between diffraction and exit point $t$ as a function of the incident angle $\omega$ and the diffraction angle 2$\theta$. Incident angle $\omega$ = -10$^\circ$ for (a) and (b), $\omega$ = 0$^\circ$ for (c) and (d) and $\omega$ = 10$^\circ$ for (e) and (f).}
\end{figure}

\Fref{distance_t} demonstrates the spatially resolved dependency of the distance between exit and diffraction point as a function of the incident angle $\omega$ and the diffraction angle 2$\theta$. As expected, $t$ generally increases with increasing distance between the diffraction point and the exit surface. The diffracted beam below the visible diagonal exits through the bottom surface of the sample, while that above the diagonal exits through the right side surface of the sample.

\begin{figure}
\centering
\includegraphics[width=0.8\textwidth]{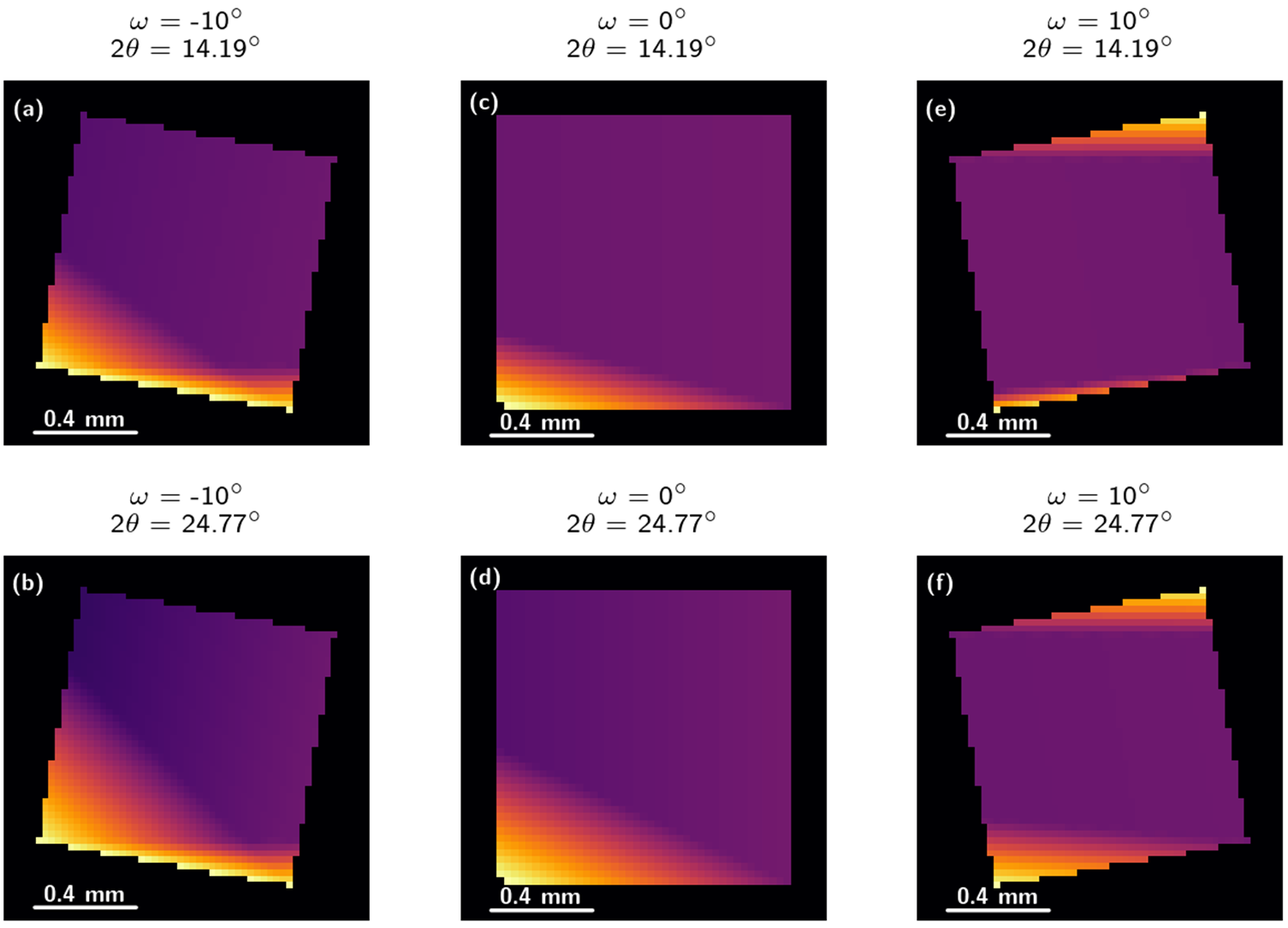}\\
\includegraphics[width=0.16\textwidth, angle =-90 ]{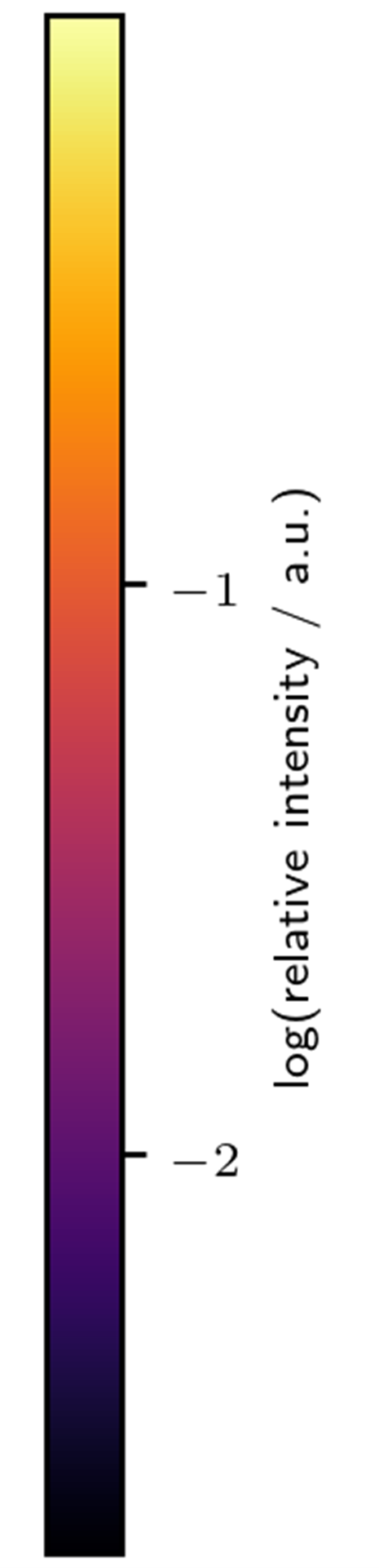}\\
\caption{\label{log10_intensity}Dependency of logarithmic intensity computed using equation \eref{beer's_law} as a function of y and incident angle $\omega$ and diffraction angle 2$\theta$. Incident angle $\omega$ = -10$^\circ$ for (a) and (b), $\omega$ = 0$^\circ$ for (c) and (d) and $\omega$ = 10$^\circ$ for (e) and (f).}
\end{figure}

\Fref{log10_intensity} demonstrates the dependency of the logarithmic relative intensity computed with equation \eref{beer's_law} as a function of the incident angle $\omega$ and the diffraction angle 2$\theta$. We observe that the diffracted intensity is two orders lower than the incident intensity, which is in agreement with a transmission through a sample, which size is four times that of the penetration depth. A noticeable asymmetry occurs with respect to the visible top and bottom parts of the sample. While the bottom part of the sample provides high diffracted intensity independent from the sample rotation $\omega$, the top part becomes available for positive incident angles, i.e., $\omega > 0^\circ$. The latter case can be explained by the fact that the X-ray beam penetrates the sample through its top surface and exits through the right hand side of the sample, which renders the total path length small in comparison to the width of the sample.

Finally, we use the 1$^\mathrm{st}$ moment of the retrieved intensities $I(x,y)$ in the incident beam direction $y$ as a representation of the average diffraction depth (i.e., statistical average over many samples). The 1$^\mathrm{st}$ moment of $I(x,y)$ is given by~\cite{moments}
\begin{equation}\label{normalisedmoment}
    M_1(y) = \ddfrac{\int x I(x,y)\, dx}{\int I(x,y)\, dx}.
\end{equation}

\begin{figure}[H]
\centering
\includegraphics[width=1\linewidth]{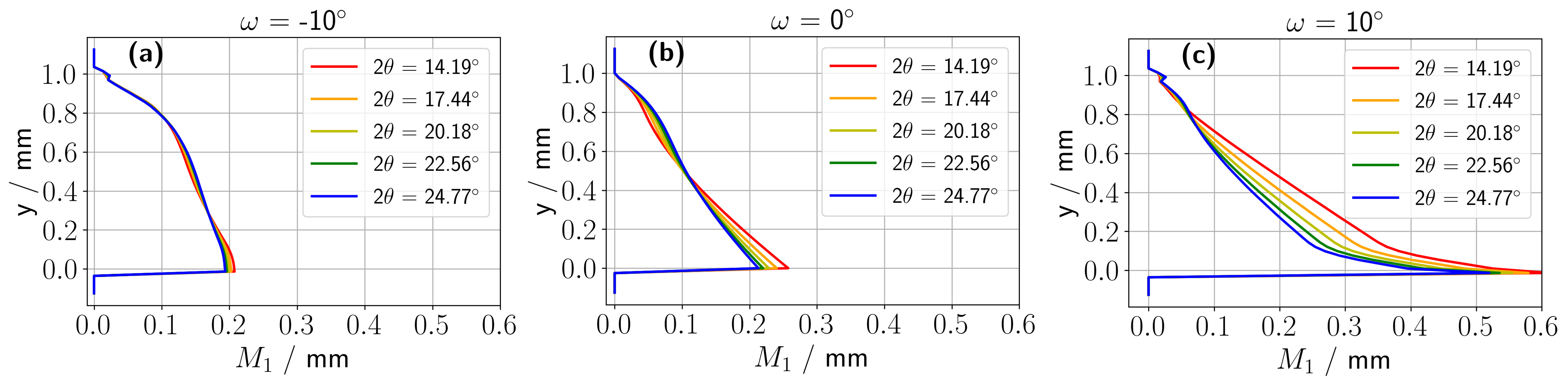}
\caption{\label{moment_firstorder}Dependency of normalised moments computed using equation \eref{normalisedmoment} as a function of y and incident angle $\omega$ (a) $\omega$ = -10$^\circ$ (b) $\omega$ = 0$^\circ$ and (c) $\omega$ = 10$^\circ$.}
\end{figure}

\Fref{moment_firstorder} demonstrates the dependency of the average diffraction depth on the incident point $y$ and sample rotation $\omega$. First, we notice that the average diffraction depth $M_1$ decreases with increasing distance between point of incident and exit surface. Here, $M_1$ varies between 0 for large $y$ and up to 0.6~mm for incidents close to the exit surface. Second and crucially, the average diffraction depth for incidents close to the exit surface varies between 0.2~mm and 0.6~mm with increasing sample rotation $\omega$. Thus, different effective sample depths can be probed by rotating the sample, which establishes a partial depth sensitivity for the described setup. Third, this can be moderately enhanced by utilizing different reflections as can be seen most noteably in \fref{moment_firstorder}(c).

\section{Conclusion} 
In conclusion, we have demonstrated that ray tracing can be utilized to numerically simulate the diffraction volume in X-ray diffraction with a focused beam. We have also shown that by rotating the sample we can probe different sample depths and thereby manipulate the partial depth sensitivity. In the future, we will compare the results of the numerical simulations with the experimental results.

\section*{References}
\bibliographystyle{iopartnum.bst}
\bibliography{iopartnum.bib}

\end{document}